\documentclass[prb, notitlepage, twocolumn, nofootinbib,superscriptaddress]{revtex4-2}

\usepackage{xcolor}
\usepackage{hyperref}
\usepackage[utf8]{inputenc}

\hypersetup{
    colorlinks = true,
    linkcolor = {orange}, 
    citecolor = {cyan},
    linkbordercolor = {white}
}

\usepackage{booktabs}
\usepackage{amsmath}
\usepackage{times}
\usepackage{graphicx}

\begin{document}
\title{ Dynamic control of octahedral rotation in perovskites by defect engineering}
\author{Jiahui Jia}
\affiliation{Catalan Institute of Nanoscience and Nanotechnology - ICN2, CSIC and BIST, Campus UAB, 08193 Bellaterra, Spain}
\affiliation{Institut de Ciència de Materials de Barcelona, ICMAB-CSIC, Campus UAB, 08193 Bellaterra, Spain}

\author{Xu He}
\email{mailhexu@gmail.com}
\affiliation{Catalan Institute of Nanoscience and Nanotechnology - ICN2, CSIC and BIST, Campus UAB, 08193 Bellaterra, Spain}
\affiliation{Institute of Condensed Matter and Nanosciences, Université Catholique de Louvain, 1348 Louvain-la-Neuve, Belgium}

\author{Arsalan Akhtar}
\affiliation{Catalan Institute of Nanoscience and Nanotechnology - ICN2, CSIC and BIST, Campus UAB, 08193 Bellaterra, Spain}

\author{Gervasi Herranz}
\affiliation{Institut de Ciència de Materials de Barcelona, ICMAB-CSIC, Campus UAB, 08193 Bellaterra, Spain}

\author{Miguel Pruneda}
\email{miguel.pruneda@icn2.cat}
\affiliation{Catalan Institute of Nanoscience and Nanotechnology - ICN2, CSIC and BIST, Campus UAB, 08193 Bellaterra, Spain}

\begin{abstract}
Engineering oxygen octahedra rotation patterns in $ABO_3$ perovskites is a powerful route to design functional materials. Here we propose a strategy that exploits point defects that create local electric dipoles and couple to the oxygen sublattice, enabling direct actuation on the rotational degrees of freedom. This approach, which relies on substituting an $A$ site with a smaller ion, paves a way to couple dynamically octahedra rotations to external electric fields. A common antisite defect, $\mathrm{Al_{La}}$ in rhombohedral LaAlO$_3$ is taken as a prototype to validate the idea, with atomistic density functional theory calculations supported with an effective lattice model to simulate the dynamics of switching of the local rotational degrees of freedom to long distances. Our simulations provide an insight of the main parameters that govern the operation of the proposed mechanism, and allow to define guidelines for screening other systems where this approach could be used for tuning the properties of the host material.

\end{abstract}
\maketitle

\section{Introduction}
The corner-sharing octahedron formed by the $B$-site transition metal ion coordinated by its six oxygen ligands in $ABO_3$ perovskites, is regarded as the fundamental functional unit in this family of materials.~\cite{rondinelli2012control,aso2014control,zhang2020unraveling,bhattacharjee2009engineering,liao2016controlled,liao2017thickness,he2016ferroelectric}. Due to the strong coupling between lattice and electronic degrees of freedom (spin, charge and orbital), it is in principle possible to fine-tune the functional properties of these materials by controlling rotations and tiltings of the $B$O$_6$ octahedra.
In the last decade great efforts have been devoted to engineering octahedral rotations by epitaxial strain or interfacial coupling in perovskite heterostructures in order to improve functionalities such as electrical conductivity, ferroelectricity, magnetism, charge and orbital ordering, superconductivity, magnetotransport, etc.~\cite{zubko2011interface,rondinelli2012control,aso2014control,kan2016tuning,liao2017thickness,liao2016controlled,he2016ferroelectric,zhang2020unraveling,bhattacharjee2009engineering} With this approach one can achieve {\it dynamic} control over the octahedra by indirect actuation over the substrate via piezoelectric coupling. However, {\it direct} coupling between octahedral rotation and external electric field is typically very weak, mostly because the corresponding atomic displacements in simple perovskites are non-polar. In more complex oxides, such as artificial superlattices~\cite{bousquet2008improper,xu2015hybrid}, or stacked perovskite double layers~\cite{benedek2011hybrid,mulder2013turning}, that show hybrid improper ferroelectricity (HIF), octahedral rotations couple to polarization, and can be dynamically tuned by an external electric field~\cite{wang2017first,xu2020highly}. However, in spite of the intensive search for HIF, these promising materials are still scarce, and the switching mechanisms can be rather complex, as the coupling involves two or more rotation modes.

On a completely different approach, dynamic control of the octahedral rotations was demonstrated by Liu {\it et al}\cite{liu2018dynamic} through the coupling with oxygen vacancies in thin WO$_3$ films. Here, ionic liquid gating is responsible for the generation of a large concentration of vacancies, which in turn results in a strong modification of the octahedral rotation. Vacancies are created by redox reactions at the oxide/gate interface, which can be reversed by changing the sign of the applied voltage. Such approach is not that different to the idea of using the chemical potential, that determines the formation of defects, as a dynamical variable similar to electric, magnetic or strain fields, that controls the properties of the host material~\cite{kalinin2013functional}. Indeed, defect engineering has become one of the more promising, yet challenging, routes to achieve new functionalities in oxides.\cite{Huang2018,choi2009role,lee2015emergence}  Furthermore, although somehow indirect in the work of Liu and collaborators, the idea of coupling lattice deformations induced by defects to an external electric field seems powerful. The concept that we introduce here goes a step beyond, by exploiting directly the local polarization induced around a point defect.

Intrinsic point defects such as cation vacancies, antisite defects, or Frenkel pairs,\cite{choi2009role,aktas2014polar} have been linked to the formation of polar nano-regions that are responsible for the ferroelectric polarization observed in KTaO$_3$ (KTO)~\cite{aktas2014polar}, or ultrathin SrTiO$_3$ (STO) films~\cite{lee2015emergence}.  In particular, first principles calculations have shown that antisite Ti defects spontaneously generate a local dipole by off-centering displacements from the Sr-site in SrTiO$_3$~\cite{choi2009role,klyukin2017effect}. Large off-centering displacements were also reported for Li and other impurities at the $A$ site in KTaO$_3$~\cite{vugmeister1990dipole}. Guided by these observations, here we propose a general design principle based on substitution of the $A$ cation with a smaller ion which potentially distorts the octahedral $B$O$_6$ tilting. Is it thus possible to bypass the competition between octahedral rotations and polar (ferroelectric) distortions in the perovskite lattice\cite{Benedek2013,Aschauer_2014,gazquez2017competition} gaining control over the rotations with external electric fields?

We focus on rhombohedral LaAlO$_3$ (rLAO)  as a prototype system for tuning of octahedral rotations by dipolar defects. Unlike SrTiO$_3$ or KTaO$_3$, pristine rLAO  has an octahedral tilt pattern at ambient conditions (a$^-$a$^-$a$^-$ in Glazer notation), due to the condensation of a soft-phonon at the $R$ point of the Brillouin zone that drives an antiferrodistortive (AFD) deformation. The antisite defect Al$_{\mathrm{ La}}$ has been previously reported for Al-rich thin films, and as for STO and KTO, the substitutional cation displaces from the ideal $A$-site position, resulting in a local dipole moment.\cite{warusawithana2013laalo} Our first principles calculations reveal that low concentration of defects induce local deformation of the octahedral tilts which are significant over a few nm length.
We also show that tilts are strongly coupled to the defect, and a relatively low energy barrier would allow switching of the rotations with external electric fields, enabling not only static but also dynamic tuning of the rotations through defect engineering. 

\begin{figure*}[t]
    \centering
    \includegraphics[width=0.6\textwidth]{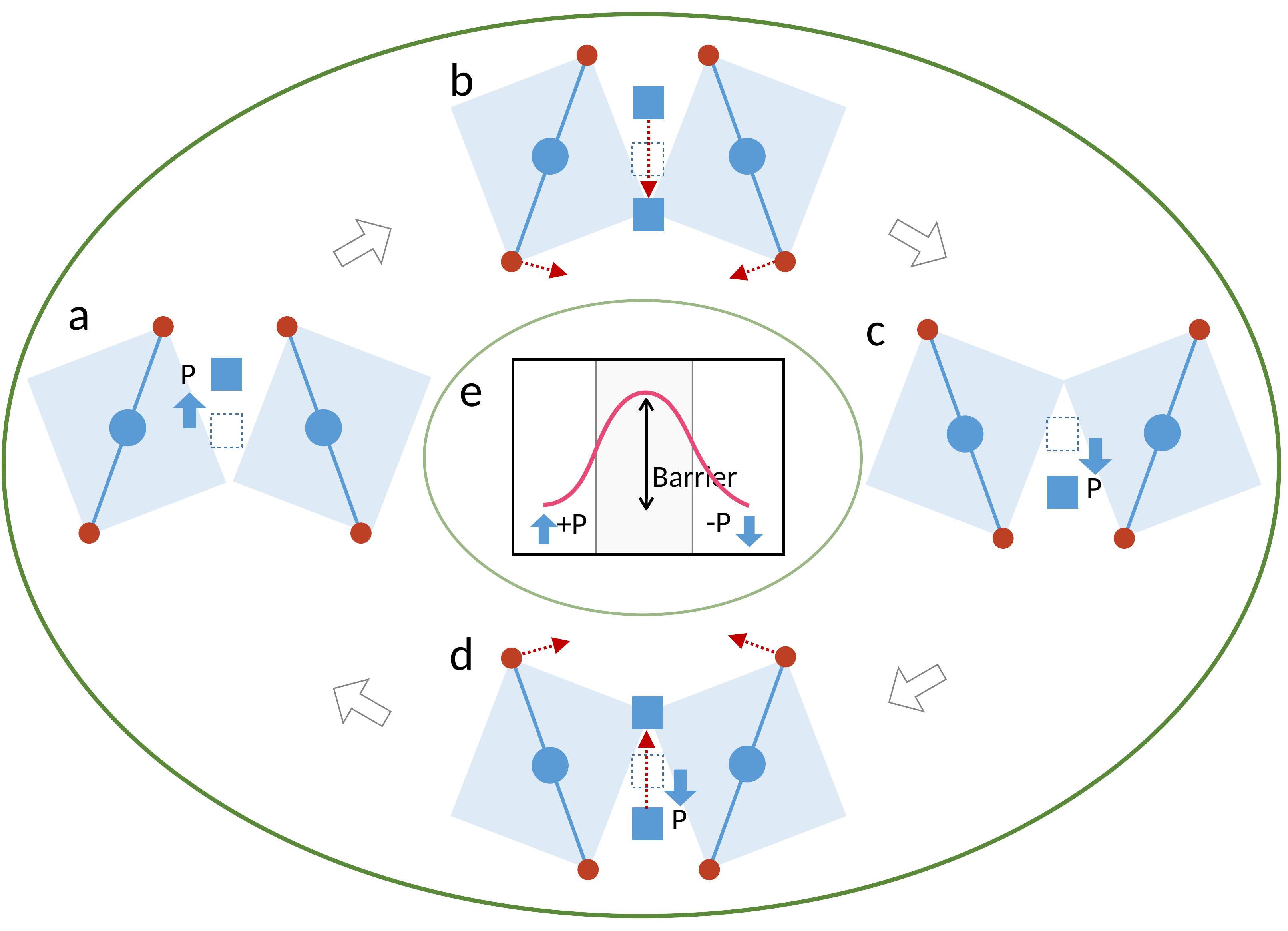}
    \caption{A schematic of the proposed mechanism to control octahedral tilts. Small red circles represent the anions at the corners of the octahedra with the $B$ cation in the center (blue circle). (a) The $A$ cation is originally placed at the dashed square. A substitutional defect at the $A$ site is denoted by filled dark blue square. This defect, which may consist of a smaller ion, such as the antisite Al$_\mathrm{La}$ in LAO, displaces preferentially towards a position close to the surrounding anions, giving rise to a local polarization $+P$. 
    This polarization couples to external electric fields, enabling the displacement of the defect to a different configuration (b). The new defect configuration is metastable, and the strong coupling to nearby O atoms modifies the rotation angle of the host lattice octahedra leading to the structure (c). Central panel (e) sketches the switching energy barrier between the two polarization states}
    \label{fig: schematic cartoon}    
\end{figure*}

The basic mechanism behind our proposed methodology is schematized in Fig. \ref{fig: schematic cartoon}. Due to the smaller ionic radius of the substitutional defect ($\mathrm{Al_{La}}$), the cation, represented as a dark blue square in the figure, will be susceptible to displacements from the original $A$ site (dashed square).  The rotation of the surrounding Al-O octahedra (in light blue) determines a preferential direction for the movement of the cation, which wants to form chemical bonds with the nearest oxygen atoms (small red circles). Hence, due to the tilt pattern in the host, the defect will displace ``up'' or ``down'' towards the closest oxygen atoms, forming the ``$+P$'' and ``$-P$'' states defined in Fig.~\ref{fig: schematic cartoon}(a) and \ref{fig: schematic cartoon}(c). The strength of the chemical bond between the cation and the oxygen atoms can also affect the rotation angle of the octahedra relative to the pristine lattice value. Furthermore, as the $\mathrm{Al_{La}}$ is a charged ion (Al$^{3+}$), an external electric field can be used to shift its position, say from ``$+P$'' state to ``$-P$'' as shown in Fig.~\ref{fig: schematic cartoon}(b). From its new position, the defect Al, which tends to form shorter chemical bonds with oxygen, will drag its neighboring oxygen atoms towards it, driving the surrounding oxygen octahedral rotation (a more realistic representation of Fig.~\ref{fig: schematic cartoon} can be viewed in Fig. S1 of the Supplementary Material~\cite{SuppMater}). We then expect a possible field-induced transition from ``$+P$'' to ``$-P$'' state, with an energy barrier that can be estimated from our first principles calculations. The switching of the octahedral rotation upon polarization switching can remain {\it local}, or propagate throughout the system. The key to the proposed mechanism is the attraction between the small substitutional ion and the oxygen atoms, and the flexibility of the octahedral rotations, which are general and thus can be extended to other materials.

The paper is organized as follows. After introducing the theoretical methodology (Sec. \ref{sec:method}), we present the results obtained in a model system used to validate our approach in Sec. \ref{sec:results}. The structural properties of the defect are discussed in Sec. \ref{sec:struct}, the energetics of the switching in Sec. \ref{sec:barrier}, and the analysis of the long range propagation of the octahedral distortions in Sec. \ref{sec:propagation}. The results and the perspectives for other systems are broadly analyzed in Sec. \ref{sec:discuss}. Finally, conclusions are outlined in Sec. \ref{sec:conclude}.

\section{Methods}
\label{sec:method}
\textbf{First principles calculations} Our DFT calculations are performed with the generalized gradient approximation (GGA) using the PBEsol exchange-correlation functional \cite{perdew2008restoring} as implemented in the {\sc Siesta} package \cite{soler2002siesta,garcia2020siesta}.  The pseudopotentials are obtained from the pseudo-dojo data set in {\sc psml} format \cite{van2018pseudodojo}, and a double-zeta-polarized (DZP) numerical atomic orbitals basis set was used to describe the electronic wave functions. A $3\times3\times4$ Monkhorst-Pack $k$-point mesh is used for the $2\sqrt{2}\times 2\sqrt{2}\times2$ supercell. Atomic positions were relaxed with a force tolerance below 0.001 eV\AA$^{-1}$.  The atomic structures were visualized with VESTA~\cite{momma2011vesta} \label{fig:forces_on_atoms}
Energy barriers between multiple stable states were obtained using the climbing image nudged elastic band (NEB) method \cite{henkelman2000climbing}. The image dependent pair
potential (IDPP) method is used to generate the initial path \cite{smidstrup2014improved}.The effect of an electric field, $\vec{\cal{E}}$, on the NEB migration path is modeled by defining the electric enthalpy functional, 
$H_E[\rho,\vec{\cal{E}}]= E\rm{_{KS}}[\rho]-\Omega \vec{P} \cdot\vec{\cal{E}}$, 
where $E\rm{_{KS}}$ is the Kohn-Sham energy functional,  $\vec{P}$ the electric polarization obtained from the modern theory of polarization, and $\Omega$ is the cell volume.

\textbf{Effective lattice model}
To follow the propagation pattern of octahedra induced by the defect, we built an effective Landau-type lattice model based on Lattice Wannier functions \cite{latticewannier} (LWF) , which has been applied in the simulation of ferroelectric properties \cite{waghmare1997ab}. A more detailed description of the model can be found in the Appendixes. With this model, we can do NVT (constant number of particle, constant volume, and constant temperature) dynamics at various temperatures by using a Berendsen thermostat \cite{berendsen1984molecular} and simulate the lattice distortion at finite temperature with the code {\sc Multibinit}~\cite{gonze2020abinit}. 

\section{Results}
\label{sec:results}
We choose rLAO as our model system because of its simplicity. The nonmagnetic La and Al allow us to eliminate possible contributions from charge, spin or orbital degrees of freedom so that we can focus on the influence of substitutional Al on octahedral rotation patterns. 
The radius of Al ion (0.535\,\AA) is much smaller than La ion (1.36\,\AA).~\cite{shannon1976revised}
Practical demonstration (and possible application) of the proposed scheme requires addressing a few aspects: What are the structural properties induced by the defect?, Is the coupling between the defect and the octahedral rotation strong enough to drive the switching of the rotation?, What is the energy barrier required to switch the defect state from ``$+P$'' to ``$-P$''?, What is the localization range of the octahedral tilt deformation induced by the defect?. In the following, we will discuss these issues in the light of our calculations. 

\begin{figure*}[t]
    \centering
    \includegraphics[width=0.8\textwidth]{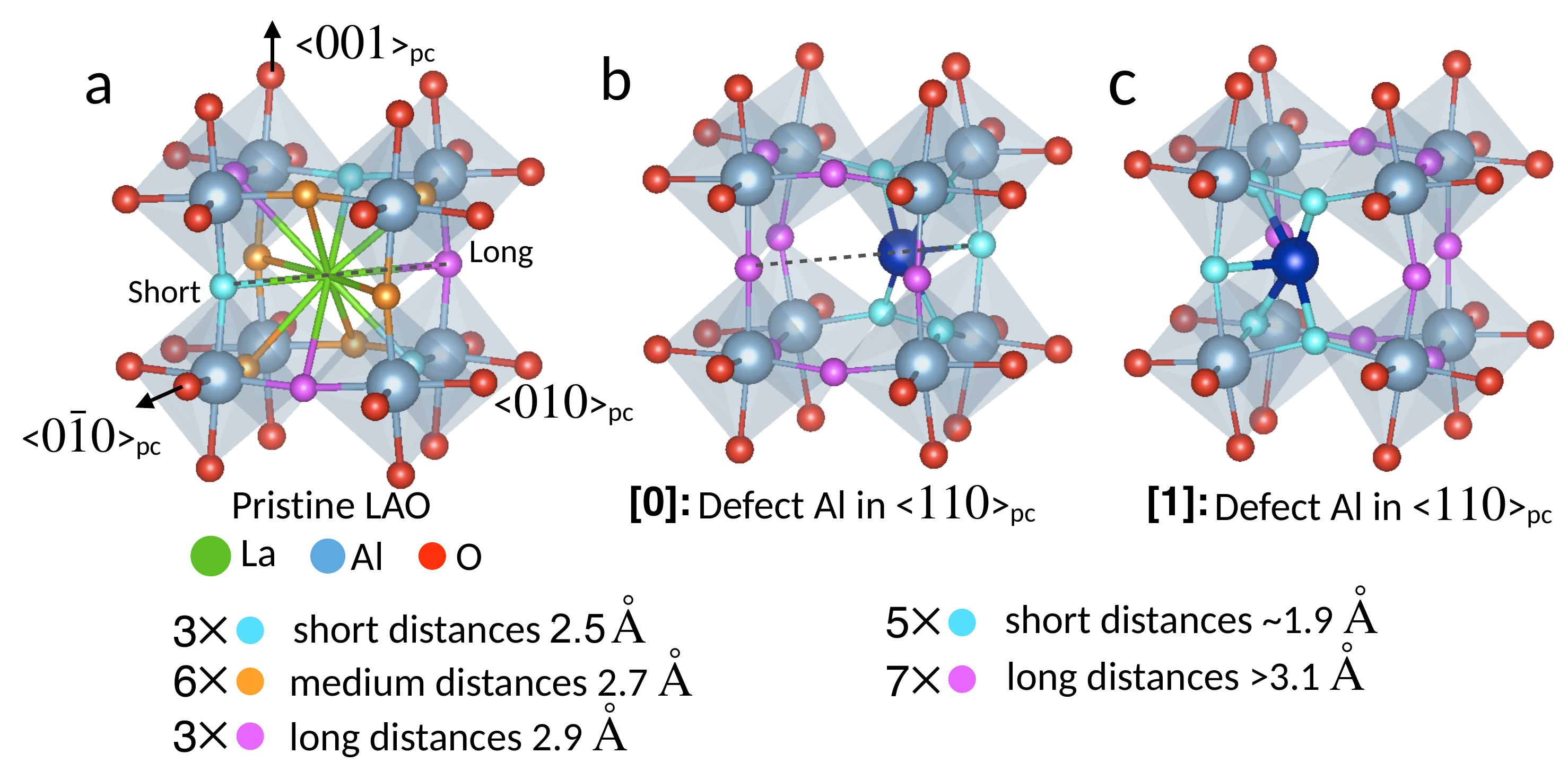}
    \caption{Local structure around the $A$ site, highlighting the eight AlO$_6$ octahedra surrounding the La site, and the different O neighbors with short, medium and long La-O distances. For local structure in (a), the substitutional Al would displace along $\langle110\rangle$, as in (b). We label this as $[0]$ site. Alternatively, a different local environment gives structure $[1]$, where the off-centering takes place along   $\langle\bar{1}\bar{1}0\rangle$, as in (c). Notice that the octahedral rotations are different for $[0]$ and $[1]$.}
    \label{fig: local structure near defect}    
\end{figure*}

\subsection{Structural characterization}\label{sec:struct}
Pristine rLAO has space group $R\bar{3}c$, and the computed  lattice parameters of the relaxed structure are $a$ = $b$ = $c$ = 5.36\,\AA\, and $\alpha$ = $\beta$ = $\gamma$ = 60.27$^\circ$, in excellent agreement with experimental lattice parameters $a$ = 5.360\,\AA\ and $\alpha$ = 60.10$^\circ$ obtained from neutron powder diffraction experiments at 15\,K \cite{howard2000neutron}.  
The Wyckoff positions of the atoms are La $2a$ $(\frac{1}{4},\frac{1}{4},\frac{1}{4})$, Al $2b$ (0,0,0), O $6e$ (0.25, 0.71, 0.79). 

We model a ${\sim}6.25\%$ defect concentration by placing one Al in a La site within a $2\sqrt{2}\times 2\sqrt{2}\times 2$ supercell. 
There are three short (${\sim}2.5$\,\AA) A--O interatomic distances along the $\langle110\rangle$, $\langle101\rangle$ and $\langle011\rangle$ directions (cyan spheres in Fig.~\ref{fig: local structure near defect}(a). In the opposite directions, the A--O distances (magenta spheres) are longer (${\sim}2.9$,\AA), and thus the defect has a preference to off-center from the $A$-site and enhance the interaction with neighboring oxygen atoms. This means that depending on the local environment (determined by the two distinct Wyckoff $2a$ sites) there will be 
two equivalent configurations,  which we identify as [0] and [1] defect states in the lattice (shown in Figures \ref{fig: local structure near defect}b and \ref{fig: local structure near defect}c, respectively), with $\mathrm{Al_{La}}$ displacing either along $\langle110\rangle$ or $\langle\bar{1}\bar{1}0\rangle$, depending on the local environment of the $A$ site (see also \ref{fig:NEB_0e}a). 
The off-centering of 1.02\,\AA\ results in a relaxation energy gain of over 4.6\,eV.

In addition to these minimal energy structures, we searched for additional configurations, by applying small displacements on Al$_\mathrm{La}$ towards other crystallographic directions. We found that there are additional local minima, in which the defect moves along $\langle1\bar{1}0\rangle$ or $\langle\bar{1}10\rangle$ [Fig.~\ref{fig:NEB_0e}(b)], both with ${\sim}0.03$ eV higher energy than the ground state in Figs.~\ref{fig:NEB_0e}(a), and (c). These displacement directions correspond to six intermediate A--O interatomic distances (${\sim}2.7$\, \AA). 
In these cases, the off-center displacement is 0.72\, \AA. Other possible structures with off-center displacements along different crystallographic directions were also obtained, all with higher energies.

\begin{table}[h]
    \centering
    \begin{tabular}{ c c }
\hline
\textbf{atoms} & \textbf{Born charge (e)}\\
\hline\\
\addlinespace[-2ex]
La & $\begin{bmatrix}  4.41 & 0.00 & 0.00 \\ -0.00 & 4.39& -0.02 \\ 0.00& -0.02& 4.40 \end{bmatrix}$ \\ \addlinespace[1.5ex]

Al & $\begin{bmatrix}  2.92& -0.08& 0.12 \\ 0.08& 2.90& -0.01 \\ -0.12& -0.01& 2.91 \end{bmatrix}$ \\ \addlinespace[1.5ex]

O & $\begin{bmatrix}  -2.44& -0.00& 0.00 \\ -0.00& -2.43& 0.01 \\ 0.01& 0.01& -2.44 \end{bmatrix}$ \\
\addlinespace[1.5ex]

$\mathrm{Al_{La}}$ in initial image & $\begin{bmatrix}  2.83& -0.00& 0.00 \\ -0.01&      3.42&     0.00 \\ -0.01 &     0.01 &     3.17 \end{bmatrix}$ \\ \addlinespace[1.5ex]

$\mathrm{Al_{La}}$ in transition image & $\begin{bmatrix}  3.31& -0.32& -0.30 \\ -0.21&      3.46&     0.06 \\ -0.21 &     0.11 &     3.25 \end{bmatrix}$ \\ \addlinespace[1.5ex]

\hline
\end{tabular}
    \caption{Born effective charges for different atoms in pristine rLAO and the substitutional defect $\mathrm{Al_{La}}$ \cite{BornCharges}}
    \label{tab:Born_charge}
\end{table}

The Glazer rotation pattern of the structure changes from $a^-a^-a^-$ in pure rLAO to $a^-a^-c^0$ in the proximity of an $\mathrm{Al_{La}}$ off-centered along $\langle110\rangle$ ($+P$ state). 
The oxygen octahedral rotation angles near this defect Al in the plane which contains the displacement vector, increase to about $\pm{7}^\circ$, which is about $\times$2 times larger than the rotation in the pristine rLAO (about $\pm{4}^\circ$). This is at the expense of a substantial reduction of the tilt rotation along the $c$ axis, which is perpendicular to the defect displacement direction. To probe the strength of the defect-tilt interaction, we took the $\mathrm{Al_{La}}$ in the structure corresponding to state ``$-P$'' as shown in the Supplemental Material~\cite{SuppMater}.
We find that the forces on the nearby oxygen atoms are strong enough to drive the rotation of the four AlO$_6$ octahedra surrounding the defect. Full lattice relaxation, even with a lower defect concentration (3.1$\%$) modeled in a  $2\sqrt{2}\times 2\sqrt{2}\times 4$ supercell, results in the defect dragging the inversion of the octahedral rotation in the whole cell. Upon switching  between the two opposite displacements [Figs.~\ref{fig:NEB_0e}(a) and \ref{fig:NEB_0e}(c)], a large dipole difference of 12.4 e$\AA$ is obtained,  which is due to the large displacement and Born effective charges~\cite{STO-compare} (reported in table \ref{tab:Born_charge}). We thus expect that application of an external field, coupled to the defect-induced dipole, can be used to switch octahedral rotations in the host lattice.

\subsection{Energy barriers}
\label{sec:barrier}
\begin{figure*}[htbp]
    \centering
    \includegraphics[width=0.9\textwidth]{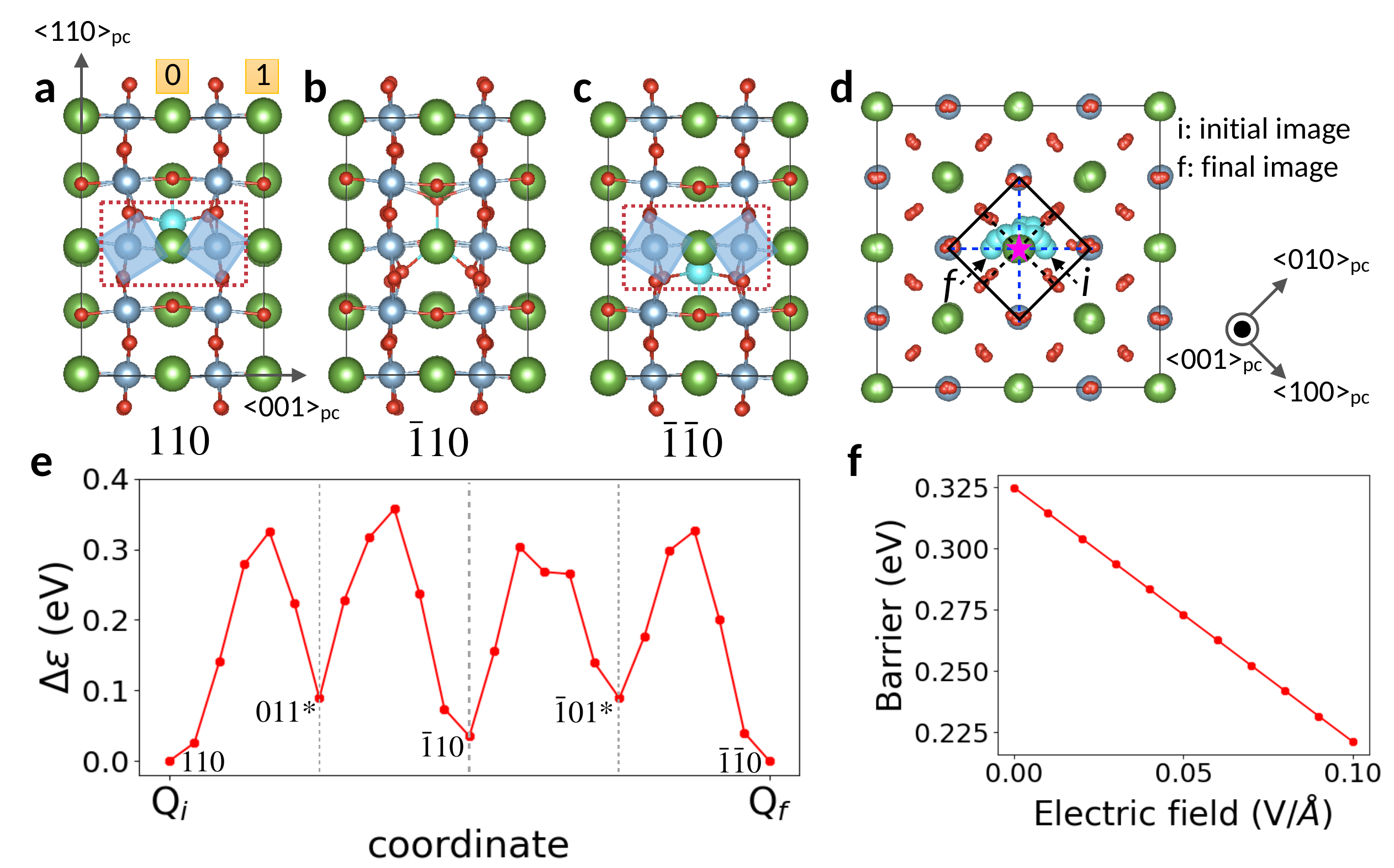}
    \caption{Relaxed DFT structures visualized with VESTA~\cite{momma2011vesta}. The supercell size is $2\sqrt{2}\times 2\sqrt{2}\times 2$, corresponding to $\mathrm{Al_{La}}$ concentration of 6.25$\%$. (a,c) are ground states with $\mathrm{Al_{La}}$ (cyan sphere) moving along $\langle110\rangle$ and $\langle\bar{1}\bar{1}0\rangle$ directions by $\sim$1.02\,\AA. Small red, large green and medium blue spheres correspond to O, La and Al atoms, and the octahedra around the defect site highlight the inversion in the rotational angles between both structures. The red dashed rectangle in the center of the supercell is a reference to the model sketched in Fig.\ref{fig: schematic cartoon} which is also revisited in \cite{SuppMater}. (b) A local minimum structure with $\mathrm{Al_{La}}$ moving along $\langle\bar{1}10\rangle$. Its energy is 0.03 eV higher than the ground states. (d) NEB transition path shown from the $\langle001\rangle$ axis, with superimposed atomic positions. Pseudocubic axes are shown together with the simulation box. (e) NEB energy barrier between (a) and (c) configurations. The off-centering directions for $\mathrm{Al_{La}}$ are indicated for each minima position. $Q_i$ and $Q_f$ are the reaction coordinates of the {\it initial} and {\it final} NEB structures, respectively. (f) Evolution of the energy barrier as a function of an electric field oriented along the $\langle110\rangle$ direction.}
    \label{fig:NEB_0e}    
\end{figure*}

In order to assess the feasibility of octahedral tilt tuning by electric field manipulation of the defect dipoles, in the following we quantify the energy barrier that needs to be overcome to move between Figs.~\ref{fig:NEB_0e}(a) and \ref{fig:NEB_0e}(c). 
We estimated the barrier by searching for a minimum energy path using the nudged elastic band (NEB) method. The na\"ive migration path in which the $\mathrm{Al_{La}}$ moves directly along the $\langle110\rangle$ direction has an energy cost of ${\sim}0.8$ eV. However, this is not the lowest possible migration path that we have identified, which takes place in two steps, and is illustrated in Fig.~\ref{fig:NEB_0e}. 
On a first stage, the $\mathrm{Al_{La}}$ defect moves along the $\langle\bar{1}00\rangle$  direction to occupy a local minimum (panel b), where the off-centering takes place along the $\langle\bar{1}10\rangle$ direction. The path to this intermediate position involves some displacement also in the $c$ direction, enabling a recovering of the AFD tilt rotation along this axis, which eases the simultaneous rotation of the octahedral lattice (the barrier for octahedral rotations in the pristine structure is ${\sim}0.05$ eV). There is a local minima along this route in which the off-centering is close to $\langle011\rangle$ (labeled in the figure as $\langle011\rangle^*$), but the $\mathrm{Al_{La}}$ bonds to only four oxygen instead of five, making this structure $\sim$0.09\,eV higher in energy than the $\langle110\rangle$. 
Finally, the $\mathrm{Al_{La}}$ is displaced towards the final $\langle\bar{1}\bar{1}0\rangle$ position (Fig.~\ref{fig:NEB_0e}(c)), following an equivalent path. 
The total migration barrier is slightly larger than the value reported for migration for $\mathrm{Ti_{Sr}}$ in SrTiO$_3$,~\cite{lee2015emergence,klyukin2017effect} but small enough to enable switching by external electric fields.

From the Born effective charges ($Z^*_i$) and the migration path coordinates ($\vec{d}_i$), we can estimate the polarization change from the initial configuration to the transition state $\Delta\vec{P}\approx Z^*_{\mathrm{Al}_{\mathrm La}}\cdot \vec{d}_{\mathrm{Al}_{\mathrm La}}+\sum_{j\ne\mathrm{Al}_{\mathrm La}}{Z^*_i\cdot\vec{d}_i}$, 
and the effect of the electric enthalpy on the migration barrier~\cite{salles2020collective}. Assuming that the Born effective charges for $j\ne\mathrm{Al_{La}}$ remain unchanged with respect to the pristine cell, the polarization work at the saddle point is given by $W^S\approx Q^{\mathrm{eff}}\cdot\vec{\textit{d}}_{\mathrm{TS}}\cdot\vec{\cal{E}}$, where $\vec{\textit{d}}_{\mathrm{TS}}$ is the integrated displacement vector of the moving atom ($\mathrm{Al_{La}}$), and $Q^{\mathrm{eff}}=\pm \left\|{{\Delta\vec{P}}}\right\|/\left\|\vec{\textit{d}}_{\mathrm{TS}}\right\|$ is an effective charge determined from the polarization variation. As shown in Table \ref{tab:Born_charge}, we have verified that the change in $Z^*$ for $\mathrm{Al_{La}}$ is less than 17\%, hence the assumption that the Born charges do not change along the computed migration path is justified.   
Taking an electric field in the direction of the initial defect polarization, we observe that the displacement of $\mathrm{Al_{La}}$ along the migration path gives a substantial change in the polarization vector and results in a migration barrier which is sensitive to the strength of the field (Fig.\ref{fig:NEB_0e}f). Hence, a sufficiently strong electric field, helped by thermal fluctuations, can be used to tune the orientation of the defect off-centering, and the chemical interactions with nearby oxygen atoms makes it possible to control the octahedral tilts in the proximity to the defect. In the following we address how the deformation of octahedral rotation induced by the defect propagates to the whole lattice.

\subsection{Propagation of the tilt deformation far from the defect}
\label{sec:propagation}

Our DFT calculations indicate that the induced octahedral tilt patterns can be reversibly switched back and forth, at least for the defect concentrations modeled with the supercell sizes used. Although concentrations above $6\%$ have been observed in experiments, it is helpful to also understand the behavior in the structure with lower concentration and up to the diluted limit. This is however, not feasible from DFT calculations due to the large computation cost. In the low-concentration limit, the interaction between defects should be neglected, and is only indirect through the octahedra rotation propagation. For this mechanism to be exploited in applications of multifunctional devices it is important to study both the dynamics of the induced rotations across the whole system, and its propagation from a single defect.
With this objective we have built an effective Landau-type lattice model based on Lattice Wannier functions~\cite{latticewannier} (see Appendix \ref{app:LWF} for details), which allows us to follow the dynamics of the local lattice distortion throughout the crystal at finite temperatures.

LWF's are close analogs to the electron Wannier functions, which form a localized basis set for describing the corresponding system. For studying the dynamics of the structure, not all of the lattice distortion modes are of equal importance. For example, the behavior of the soft modes is often enough for understanding structural phase transitions. Therefore the dynamics of the system could be studied with a Hamiltonian with a minimal basis set which could well describe the subspace of distortions of interest. The LWF's form such basis. The phonon branches that correspond to the octahedral rotations can be used to construct the minimal basis Hamiltonian that allows to study the dynamics of such octahedral rotations. 
Each LWF (labeled $i$) is a group of atomic displacements (the amplitude of which is $\tau_{iu}$ for the atomic displacement $u$) within a certain local range. The amplitude of the LWF is then defined as $W_i^2 = \sum_u\tau_{iu}^2 $. The octahedral rotation in pristine rLAO relative to the cubic structure determines the amplitude $W^{(0)}$ of the LWF's in bulk.

\begin{figure*}[t]
    \centering
    \includegraphics[width=0.55\textwidth]{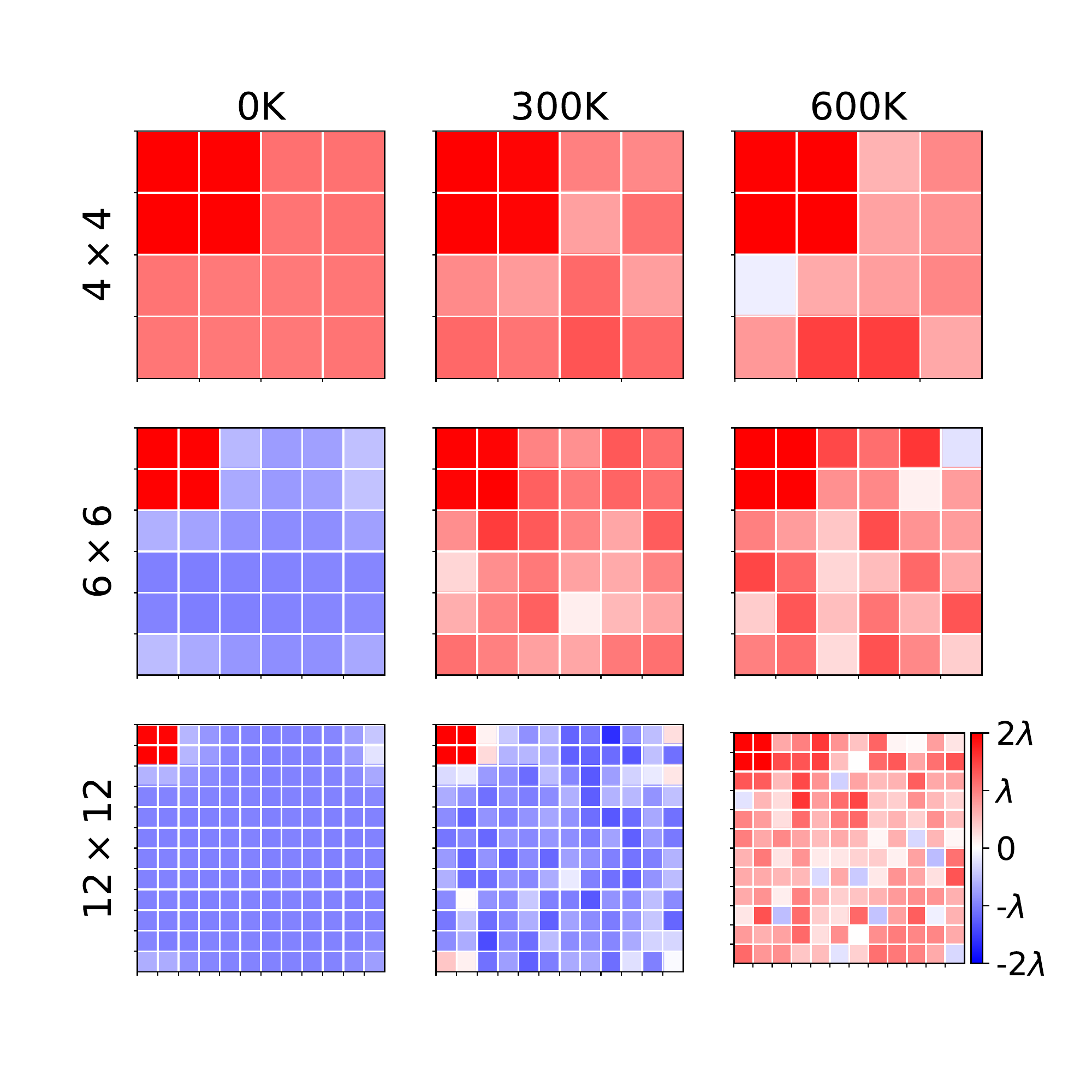}
    
    \caption{Map of the staggered amplitudes of the LWF's after 3.6\,ps total simulation time (time step 0.1\,fs) in various supercell sizes ($4\times 4 \times 4$, $6\times 6 \times 6$, and $12\times 12 \times 12$ from top to bottom corresponding to defect concentrations of 1.56\%, 0.46\%, and 0.06\% respectively), and simulation temperatures (0.1K, 300K, and 600K from left to right). The fixed LWF amplitudes at the top left corner of each supercell are double those of the pristine structure ($\lambda$=2). The reversed octahedra are shown in red, while the non-reversed are in blue.}
    \label{fig:domain2.0}
\end{figure*}

As mentioned above, the defect induces an increase by a factor close to 2 for the tilt angle.  
Therefore, in our LWF model, we set the amplitudes for the eight octahedra 
($2\times 2 \times 2$ cell) 
closest to the defect to $\lambda\times\,W^{(0)}$, with $\lambda=2$.
Thus, by switching locally the amplitude of the LWFs around the defect, we monitor how that fixed perturbation propagates in the system for a given temperature.  
Note that due to the corner-sharing connectivity of perovskites, the rotation of the octahedra along a particular axis predominantly propagates in the plane perpendicular to that axis. 
We follow the evolution of the ``staggered'' amplitude of the LWF's, $W_i^S$, defined as:  
$W_i^S=W_i \exp(2\pi i\, \textbf{\textit{q}} \cdot \textbf{\textit{R}}_i)$
where $W_i$ is the amplitude of the LWF with index $i$, $\textbf{\textit{R}}_i$ is the corresponding cell vector, and $\textbf{q}$=$(\frac{1}{2},\frac{1}{2},\frac{1}{2})$ is the wave vector of the unstable phonon that gives the bulk tilt pattern. 
Results obtained for different defect concentrations (1.56\%, 0.46\%, and 0.06\%) and sampling temperatures from 0 to 600 K (below the transition temperature to the cubic phase at ${\sim}800$ K, Fig.~\ref{fig:LWFTc}), are summarized in Fig.~\ref{fig:domain2.0}. The image shows the final configuration after 3.6 ps, where the rotations of the four squares in the top left corner (in red) were switched with respect to the initial octahedral rotation configuration of the system (negative $W^S$, in blue).

Our simulations reveal that for very small concentration of defects (lower panels in Fig. \ref{fig:domain2.0}) the propagation of the rotation switching to the whole lattice is hindered. However, for larger concentrations (0.46\% and above) thermal fluctuations seem to be enough to enable the inversion of the whole octahedral rotation.  This happens if the strength of the defect-octahedra rotation wins over the softness of the phonon mode: in other words, the lattice is not able to switch back the defect position (and thus the local octahedral rotation).
Similar conclusions are obtained if the local deformation induced by the defect is weaker (smaller rotations in the proximity of the defect, modeled with $\lambda=1$). Although in this case the defect concentration or the temperature have to be increased to ensure the propagation of the switch to the whole lattice (Fig.S2, in SI) 
the effect seems to remain possible for temperatures well below the transition temperature.  Note that samples with La/Al ratios below 0.97, which have been related to high concentration of defects (likely antisites),~\cite{warusawithana2013laalo,qiao2011epitaxial} are not uncommon in the laboratory and can be conveniently engineered in samples grown by Pulsed Laser Deposition (PLD)\cite{ohtomo2004high,qiao2011epitaxial} or Molecular Beam Epitaxy (MBE)\cite{segal2009x}. 

\section{Discussion}
\label{sec:discuss}
Our test bed system to demonstrate the feasibility of the proposed approach has been the antisite Al$_{\mathrm {La}}$ defect in LAO. However, substitutional defects, where the $A$ site is occupied by a smaller ion, are ubiquitous in perovskite structures, and have been reported not only in LAO, STO, and KTO, but in many other systems such as TbMnO$_3$ ($\mathrm{Mn_{Tb}}$) \cite{farokhipoor2014artificial}, BaSnO$_3$ ($\mathrm{Sn_{Ba}}$) \cite{scanlon2013defect,kc2019first,combs2020stoichiometry}, SrSnO$_3$ ($\mathrm{Sn_{Sr}}$) \cite{kc2019first}, YAlO$_3$, LuAlO$_3$ ($\mathrm{Al_Y}$, $\mathrm{Al_{Lu}}$) \cite{singh2007antisite}, LaScO$_3$ \cite{islam2004doping}, LuFeO$_3$ ($\mathrm{Sc_{Lu}}$) \cite{deng2021critical}, or NaTaO$_3$\cite{choi2008first,sudrajat2020water}. Substitutional defects have also great influence on the properties of halide perovskites such as FAPbI$_3$\
(FA=HC(NH$_2$)$^{2+}$)~\cite{saidaminov2018suppression}.  Hence, our strategy is likely to be suitable for a broad variety of systems. 

One can expect that a smaller cation in the $A$ site can produce a distortion of the perovskite lattice by a contraction of the lattice and/or a tilting of the $B$O$_6$ octahedral sublattice. Both effects will have consequences on the electronic properties of the material. However, contraction of the lattice is expected to be localized around the defect, while the distortion of the octahedral rotations can propagate to longer distances, as our calculations have shown. Hence, even a small concentration of defects can have global effects on the system. Furthermore, actuation by electric fields on the local dipole open the possibility for dynamic control of the electronic structure. Keep in mind that the increase in the octahedral tilt angle results in poorer overlaps of the orbitals for the metal at the $B$ site, which will result in narrower conduction bands, weaker magnetic interactions, etc. 
Indeed, in our example system changes in the tilt angle by a factor 2, as obtained in our defective supercell, can give a band gap increase by up to a 30\% (at the DFT level).
 
What materials can be identified as potential candidates for defect-engineering the octahedral rotation, and consequently the electronic properties? Our first-principles calculations, and the effective Landau-type lattice model can help to determine which parameters control the propagation of the local octahedral rotation to long distances. The presence of a soft-phonon in the high-symmetry phase of the perovskite ensures the rotation of the octahedra from the untilted cubic phase, and the presence of a potential double well. Although the depth of the well is not directly determined by the energy of the soft phonon, a harder soft phonon points in the direction of a shallower potential well, which would be easier to reverse (Appendix \ref{app:factors}). In addition, the localization of the LWFs (related to the rotation of an octahedron following the soft phonon) gives an idea of the coupling with nearby LWFs. This localization is related to the curvature of the phonon dispersion $d^2E/dq^2$, so that a flat dispersion gives more localized LWFs, weaker interaction with neighbors, and thus octahedral rotations easier to reverse. A fast screening of phonons in the \texttt{MaterialsProject} data set~\cite{jain2013commentary,phonondb} for simple perovskites identifies NaTaO$_3$, SrSnO$_3$, KCaF$_3$, RbCaF$_3$, PrAlO$_3$, SrHfO$_3$, and NaNbO$_3$ as promising candidates to propagate local tilt distortions to long distances.
It remains to be checked if (and which) substitutional defects show off-centering and induce local dipole moments in these systems. It is worth stressing that the database mentioned above does not include many transition metal oxides which host interesting spin, charge and orbital properties which can be tuned by the octahedral rotations.

One example in which defects have been used to tune the electronic properties via deformation of the octahedral rotations is La-doping in NaTaO$_3$, where a gradient in the spatial distribution of dopants has been shown to induce a bending in the conduction band that favors electron-hole separation for photocatalysis~\cite{sudrajat2020water}. This example poses an additional ingredient in the discussion, as the dopant has a different ionic charge than the atom it substitutes (La$^{3+}$ vs Na$^+$). This is not the case in the system that we studied here (Al$^{3+}$ vs La$^{3+}$) and hence the effect on the physical properties is purely based on the distortion on the octahedral rotations (and not other effects, such as valence state, or magnetic moments, for example). 
 However, we did verify the behavior of the defect in different charge states (see details in the Supplemental Material~\cite{SuppMater}, including Ref. ~\cite{FNV} therein). Al$_{\mathrm{La}}$ has an in-gap defect level which can trap up to two electrons.  If the level is fully occupied the ground-state configuration of the system lacks the off-centering and the local dipole is suppressed. The question of whether electronic photoexcitations could be used to activate the defect polarization, or its switching, is beyond the scope of this work.

Here we have only considered an ordered distribution of defects, with relatively low concentrations, which give local dipoles (polar nanoregions, PNR). These PNRs would be randomly distributed, along the $[110]$, $[101]$ or $[011]$ directions, and the total net polarization would be canceled. Nevertheless, there is a long history of research on the switching properties of PNR in the so called {\it relaxor ferroelectrics}. Application of an external electric field can shift the direction of the dipole (by migration of the off-centered defect to a different site), as we have investigated.
However, more complex scenarios can be envisioned where different defects interact with each other. Antisite Ti$_{\mathrm{Sr}}$ defects in SrTiO$_3$ have been shown to bound to oxygen vacancies, affecting the dipole of the complex defect, and increasing the activation barrier for polarization switching~\cite{choi2009role,klyukin2017effect}. A similar effect can be expected for Al$_{\mathrm{La}}$ in LaAlO$_3$. Furthermore, it is possible that different substitutional defects occupy the two distinct $A$ sites, which would give opposite polarizations ($+P$ and $-P$), and a further increase in the switching barrier. One possible route to mitigate this issue requires breaking the symmetry of the $A$-site, for example using double perovskites ($A/A'$)$B_2$O$_6$, so that the substitution can be favored in one particular site, giving a preferential polarization direction. These scenarios require further investigation. 

In perovskites the structural, electronic and magnetic properties are often strongly coupled with the oxygen octahedra rotations. 
Some other lattice distortion  modes, like the Jahn-Teller distortion in the manganites, titanates, and vanadates, or the breathing distortion in the nickelates and ferrites, which are related to the metal-insulator transition, can be modulated by the octahedral rotation through phonon-phonon coupling \cite{lee2013strong,balachandran2013interplay,varignon2015coupling,mercy2017structurally,zhang2018from}. Moreover, the octahedra rotation modulates the overlap of the orbitals on the $B$-site and the oxygen $p$ orbitals, thus tuning the band width which is essential in the Mott-Hubbard physics. This also affects the superexchange and double-exchange in magnetic materials which follow the Goodenough-Kanamori rules\cite{anderson1950antiferromagnetism,goodenough1955theory,goodenough1958interpretation,kanamori1959superexchange}. The octahedra rotation pattern is also decisive for the Dzyloshinskii-Moriya interaction (DMI), which is directly related to the $B$-O-$B$ bond angle\cite{dzyaloshinsky1958thermodynamic,moriya1960anisotropic}. The DMI leads to weak ferromagnetism\cite{bousquet2016non} and its modulation by electric field leads to new possibilities for magneto-electric coupling\cite{benedek2011hybrid}.   Therefore, dynamical tuning of the octahedral rotation by polar defects can bridge the electronic/structural/magnetic properties, which can not be easily achieved in most pristine structures.  Since the method proposed here is likely to be general for perovskites, we expect it will trigger further investigation for a broad range of possible applications.

\section{Conclusion}
\label{sec:conclude}
In conclusion, we propose a method for tuning octahedral rotation based on defect engineering, which can be widely extrapolated to $AB$O$_3$ oxides with perovskite-like structure. The dynamic control can be achieved by coupling electric potentials to the local dipole moment induced by the off-centering of the substitutional defect. The concept has been demonstrated by DFT calculations on a common antisite defect in LaAlO$_3$, where the strong coupling to the oxygen sublattice, and the low barrier for switching the position of the defect enables dynamic control of the physical properties that are sensitive to $B$-O bonds, including electronic, magnetic and optical characteristics.  A simple model based on lattice wannier functions allows to draw some guiding principles for selecting materials where the long-range propagation of local octahedral rotations is possible, with moderate defect concentrations. This method can be combined with other approaches to tune the octahedral rotations in multifunctional materials, such as strain engineering or interfacial coupling, opening new exciting opportunities for future research.

\section*{Acknowledgements}
JJ, XH, AA, and MP acknowledge financial support from PGC2018-096955-B-C43, funded by MCIN/AEI/ 10.13039/501100011033 and by “ERDF A way of making Europe”, and Generalitat de Catalunya (Grant No. 2017SGR1506). We also acknowledge the European Union MaX Center of Excellence (EU-H2020 Grant No. 824143), and INTERSECT (H2020-NMBP-TO-IND project GA n. 814487). ICN2 is supported by the Severo Ochoa program from Spanish MINECO (Grant No. SEV- 2017-0706) and the CERCA Program of Generalitat de Catalunya. JJ and GH acknowledge the financial  support from PID2020-118479RB-I00 and Severo Ochoa FUNFUTURE (CEX2019-000917-S) projects of the Spanish Ministry of Science and Innovation (MCIN/AEI/10.13039/501100011033). XH acknowledges financial support from F.R.S.-FNRS through the PDR Grants PROMOSPAN (T.0107.20). JJ acknowledges financial support from China Scholarship Council (CSC) with no. 201904910557.

\appendix

\section{Lattice Wannier function model}
\label{app:LWF}
Using large supercells in DFT simulations is computationally challenging. Instead, we have used an effective Laudau-type lattice model based on lattice wannier functions (LWF) to study the long-range dynamics of octahedral rotation distortions at finite temperatures.
This method has been previously applied to study ferroelectric phase transitions\cite{waghmare1997ab}.   In analogy to the electronic Wannier functions, LWF are a set of localized functions in real space, which can be seen as a Fourier transformation from a subset of the phonon Bloch functions.
The phonon subset (in this case, octahedral rotations) can be selected from a defined energy range in the vibrational spectra. The selected LWF's form an ideal minimal basis set that describes the local rotation distortions. In the harmonic approximation the model preserves the unstable phonons. However, anharmonic terms can be easily fitted thanks to the reduced number of degree of freedom.

Here, the LWF's and the corresponding lattice Hamiltonian are constructed from the selected columns of density matrix in $k$-space (SCDM-$k$) method \cite{damle2015compressed,damle2017scdm,damle2018disentanglement} as implemented in the
{\sc banddownfolder} \cite{banddownfolder} package. Although the SCDM-$k$ has been introduced for electronic Wannier functions, it can be extended to lattice Wannier functions. The basic idea behind this method is that the columns of the density matrix $\mathbf{\rho}$ are usually localized, and thus could be used as Wannier functions.
The columns with indices $C$, denoted as $\mathbf{\rho}_C$, are selected to best approximately span the columns of $\mathbf{\rho}$, with the rank-revealing QR decomposition, where Q is an orthogonal matrix  and  R is an upper triangular matrix.
The density matrix is defined as $\mathbf{\rho} = \mathbf{\Psi F \Psi^\dagger}$, where $\mathbf{\Psi}$ are the wave functions, and $F$ the occupations.  For electrons, $F$ is determined by the Fermi distribution function, although it can be generalized to any function that represents the weights of eigenmodes. Among the most commonly used functions are the Gaussian which select the bands around a value within certain width, the unity function which include all the bands equally, and the Fermi function which selects the bands below a value. 
Smooth functions are often used, e.g., Fermi function with a large smearing.  For lattice Wannier functions, $\mathbf{\Psi}$ are the phonon eigenmodes.

In the crystal structure, the density matrix can be defined for each wave vector $q$ (for electrons, $k$ is often used, thus, the method is named as SCDM-$k$.),  $\mathbf{\rho_q} = \mathbf{\Psi_q F \Psi_q^\dagger}$. The same method for selection of the rows (the indices of rows is denoted as $C$, which has the size of the number of Wannier functions.) can be applied to one so-called anchor point $q$. Then, the same columns are selected for the rest of $q$ points, denoted as $\mathbf{\rho}_C (\vec{q})$. By transforming it from the reciprocal space to real space, the LWFs can be  calculated as the columns of $\mathbf{\omega}(\vec{R})= \int_{q\in {\rm BZ}} \mathbf{\rho}_C(\vec{q}) e^{-i\vec{q}\cdot{\vec{R}}}d q$, where $\vec{R}$ is the cell vector. 

\begin{figure}[htbp]
  \centering
  \includegraphics[width=0.45\textwidth]{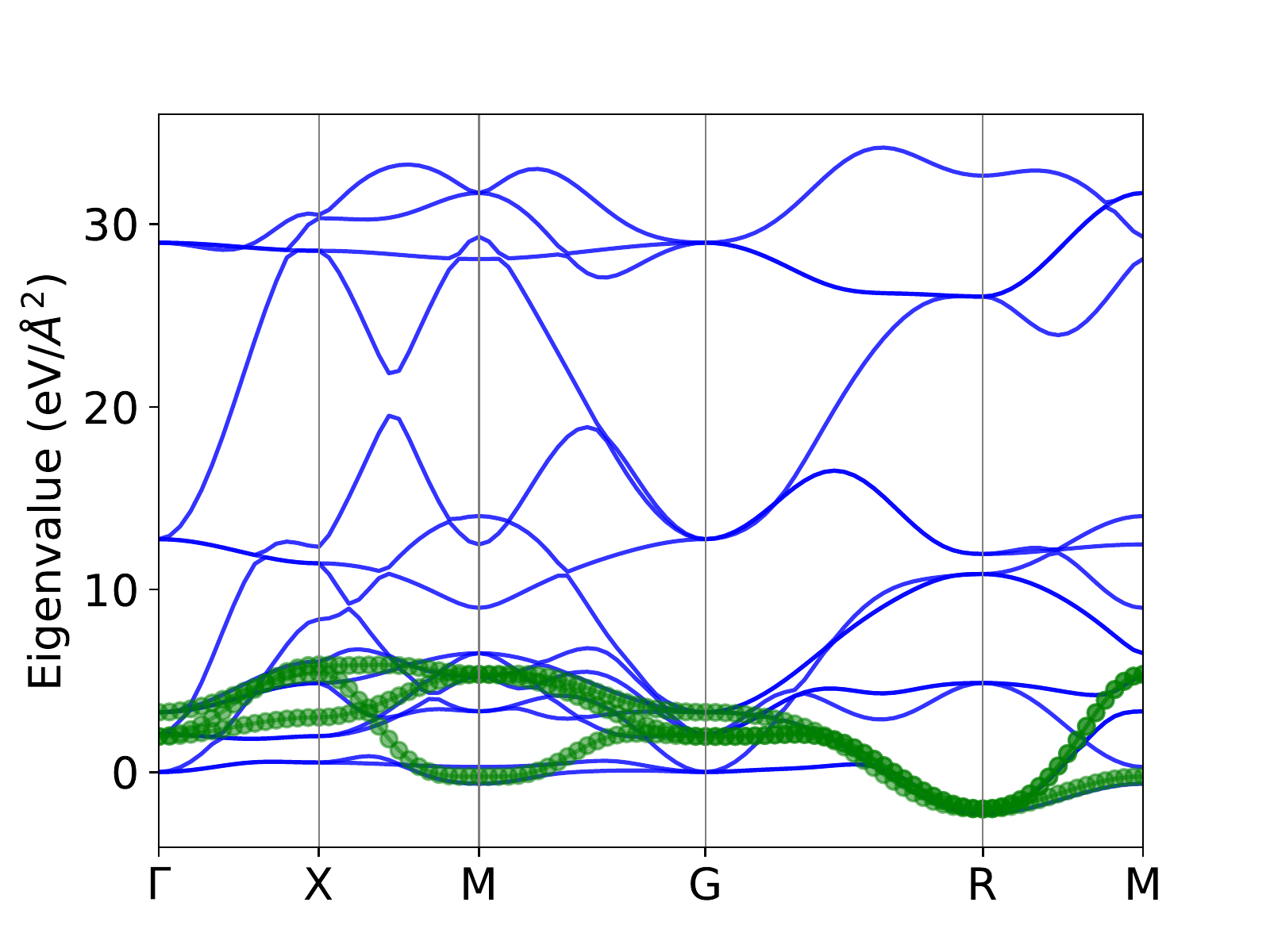}
  \caption{The eigenvalues of the IFC\'s (blue) and the LWF (green) band structure. }
  \label{fig:lwf}
\end{figure}

To build the LWF's, we take the interatomic force constants (IFC) of cubic LAO as an input. A Fermi function with Fermi energy at 0 eV and smearing of 0.3 eV, is taken as a weight function to make sure that the negative eigenmodes are well represented by the LWF. The large smearing makes the fitting smoother.  The description of the dispersion curves using LWFs is in good agreement with the unstable phonon branches obtained directly from the IFC (Fig. ~\ref{fig:lwf}), especially in the proximity to the unstable modes at the $R$-point, $\vec{q}$=($\frac{1}{2}$,$\frac{1}{2}$,$\frac{1}{2}$), responsible for the anti-phase octahedral rotations that describe the rhombohedral $a^-a^-a^-$ phase.

The energy is written as a Landau-like expansion with $W_i$, the amplitude of the
LWF's, as order parameters:
\begin{equation}
  \label{eq:lwfham}
  E= \sum_{ij} D_{ij}W_{i}W_{j} + \sum_i C_4 W_{i}^4 \,,
\end{equation}
where the first term gives the harmonic term, and the second term includes an on-site higher order contribution. $D_{ij}$ is calculated from construction of the LWF's,  while $C_4$ is obtained from a fitting to the DFT energies for various random $W_i$. The second order derivative of the energy with respect to the octahedral rotation is negative at $W_i = 0$, which gives the phonon instablity and is compensated by the positive fourth-order term as the amplitude increases. This minimalist model, including the lowest order inter-LWF interaction, and the second- and fourth-order intra-LWF interaction, captures the essential physics: the correlation between the neighboring rotations and the instability of the rotations. For a more accurate study, the stable phonon
branches and the interactions between them, the inter-LWF anharmonic terms, or the coupling with strains fields, should be considered. However, in this work, we focus on the understanding of the essential physics and avoid these complexities. 

With this model, we run NVT dynamics at various temperatures with a modest computational cost. The initial state configuration is set to the pristine structure, except for the close neighborhood to the defect, which is fixed to a reversed state configuration (as shown in the upper left $2\times 2$ corner cells in Fig.~\ref{fig:domain2.0} and Fig. S2). We then run NVT lattice dynamics, with a time step of 0.1\,fs for a total run of 3.6\,ps, using different supercell sizes to check whether the local inversion can propagate throughout the system.

We multiply the amplitude of the LWF's by the phase factor of the cell it is centered at to get the amplitude of the phonon at $\vec{q}$=($\frac{1}{2}$,$\frac{1}{2}$,$\frac{1}{2}$). 
Thus we define the "staggered" amplitude for each LWF:  $W_i^S=W_i \exp( i \vec{q} \cdot \vec{R_i})$, where $\vec{R}_i$ is the cell vector for the corresponding LWF. 

\begin{figure}
  \centering
 \includegraphics[width=0.45\textwidth]{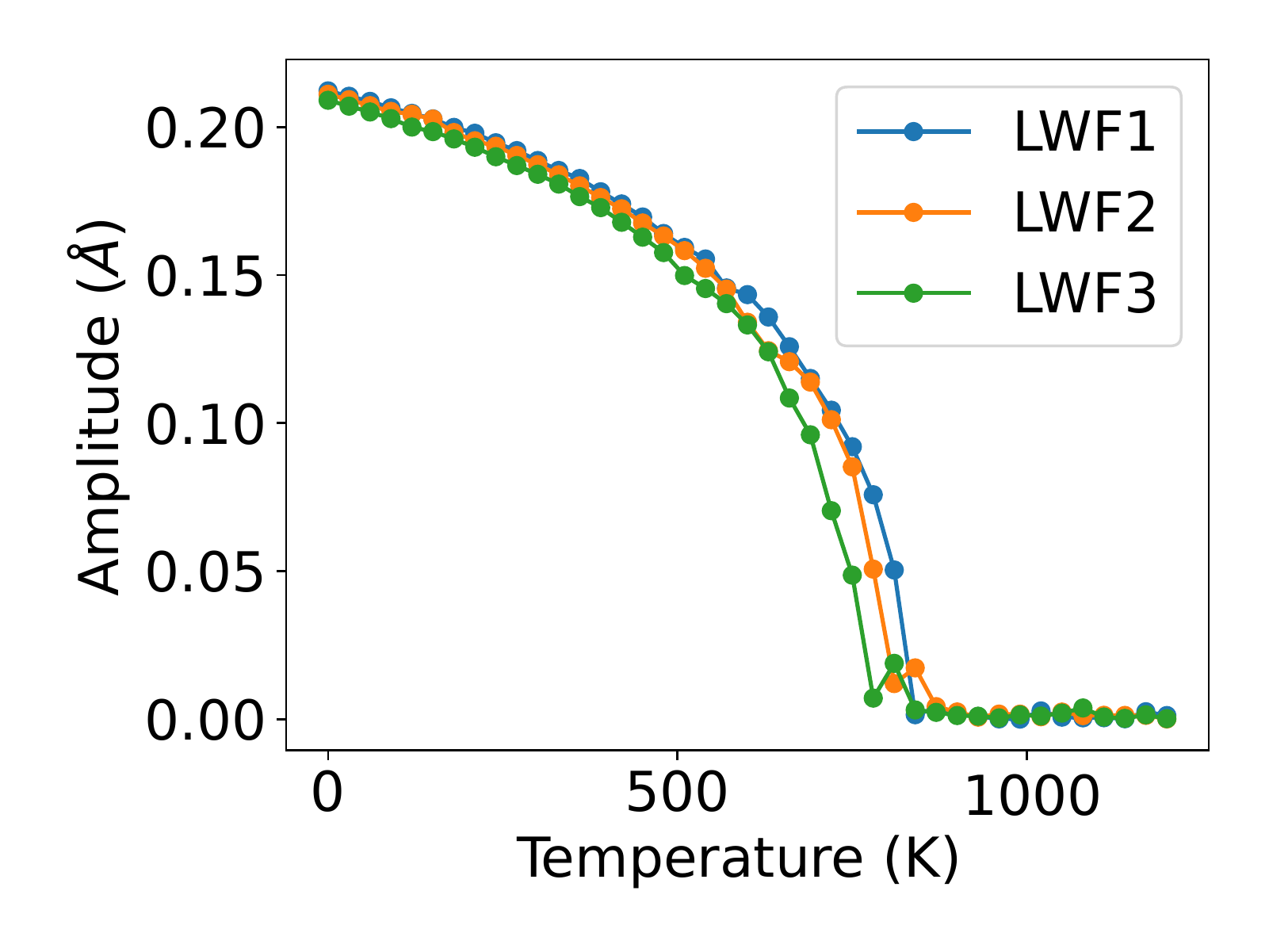}
 \caption{The average amplitude of the LWF's ($\left< W \right >$) as functions of temperature. The three lines represent the three LWF's in each unit cell. \label{fig:LWFTc}}
\end{figure}

In each unit cell, there are three LWF's corresponding to the three phonon branches at $R$. We can calculate the averaged value $\left< W^S \right > = \frac{1}{N} \sum_{i=1}^{N} W_i^S$ for a fixed temperature, as shown in Fig. ~\ref{fig:LWFTc}. The averaged amplitudes of the three LWF's are about the same at low temperatures ($R\bar{3}c$ phase), and decreases with increasing temperatures until it eventually falls to 0 for the cubic phase transition point at about 800 K, in agreement with the experimental $T_C$ of 800 K\cite{lehnert2000powder}. 
Although this remarkable agreement is to some extent a coincidence, it shows the model captures quite well the essential physics for the structural transitions.  

\section{Factors affecting the propagation of the octahedral rotation deformation}
\label{app:factors}
\begin{figure}[h]
    \centering
    \includegraphics[width=0.45\textwidth]{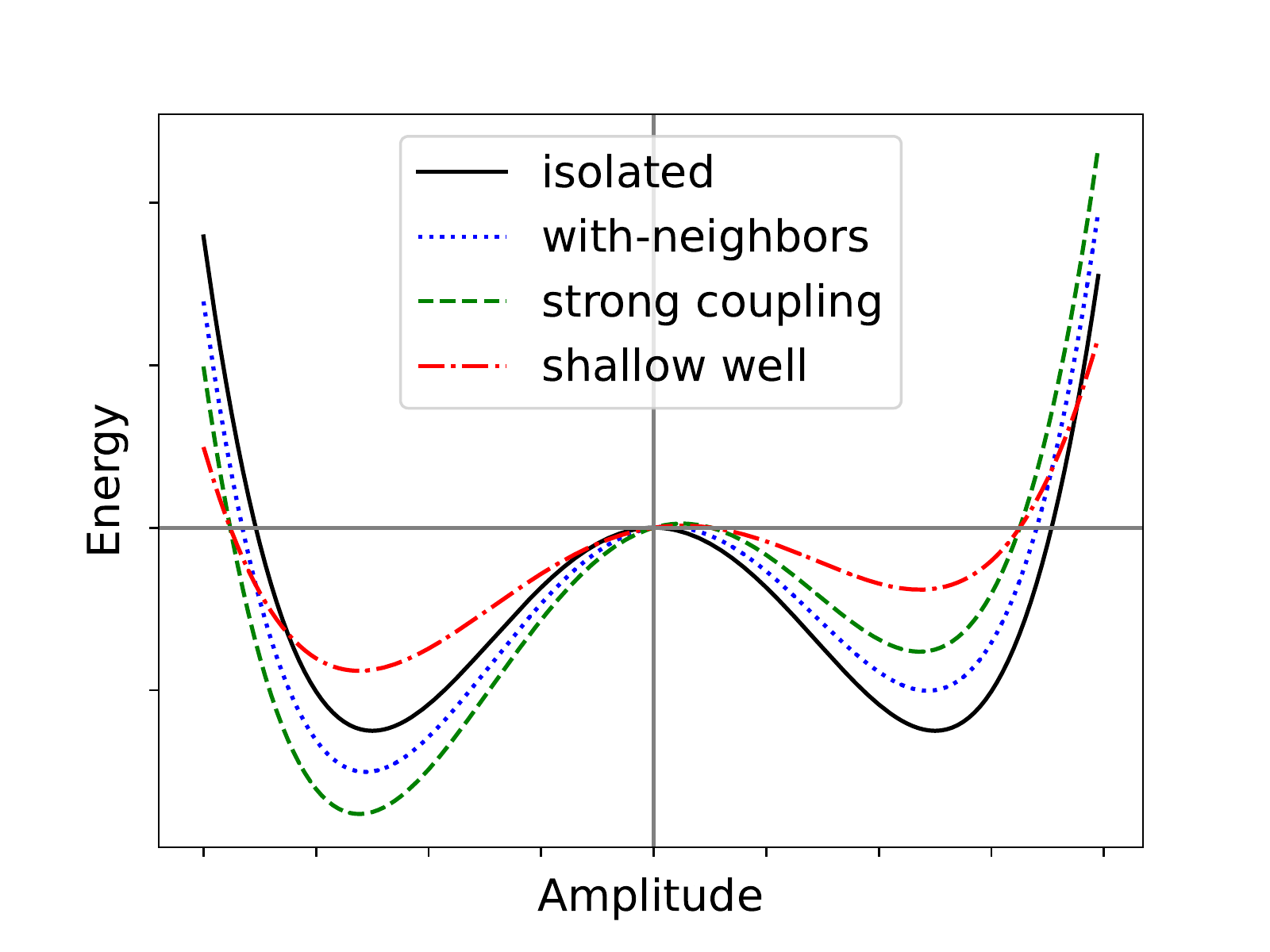}
    \caption{The schematic view of the potential energy surface by varying the amplitude of the rotation. Solid black: the symmetric potential well from the "self-interaction" terms. Dotted blue: the mean field of the interaction with neighboring sites are added. Compared with the blue curve, the green dashed curve has a stronger coupling, and the red (dashed-doted) curve has a shallower double well. }
    \label{fig:doublewell}
\end{figure}

In order to understand the factors that influence the propagation of the rotation pattern, we consider the behavior of each local octahedral rotation in a mean-field approximation, rewriting the LWF Hamiltonian:
\begin{equation}
    H= A_2 \Phi^2 + A_4\Phi^4 + B n_N \Phi \left< \Phi_N  \right> \,,
\end{equation}
where $\Phi$ represents the staggered local distortion amplitude.  The first two terms give the potential double well, which gives two stable configurations with positive or negative octahedral rotations. The $B$ in the third term, is the coupling strength that couples $\Phi$ with the ``mean value'' of the neighboring sites $\left< \Phi_N  \right> $, and $n_N$ is the number of neighbors. 
The energy landscape is shown in Fig. ~\ref{fig:doublewell}. If the coupling factor $B$ is weak, the two minima are equivalent (black solid line in the figure). However, a stronger coupling breaks the $\pm{\Phi} $ symmetry and favors one particular configuration, affecting also the energy barrier between them (blue dotted line).

Let us assume that the initial state is at $\Phi<0$. For the cells close to the defect, $\left<\Phi\right> $ is the result of the competition between the cells constrained by the defect (which are more strongly bound than the double-well minima) and the free-moving ones with negative amplitude. Initially, $\left< \Phi_N  \right> <0 $ as the constrained cells are out-numbered by the free ones. 
The probability for the local rotation to switch is related to the barrier of the transition.
The depth of the potential well (determined by the first two terms), and the coupling strength $B$ (third term in the equation) essentially defines the shape of the energy barrier. These parameters can be linked to the phonon dispersion curves, and can be used to pre-screen promising materials for future investigations.
First, a soft-phonon mode must exist, to enable the double-well structure and expontaneous rotation of the octahedra. The harder this phonon, the shallower the potential well is, making it easier for the octahedra to reverse upon a perturbation (red line in the figure).
In addition, the curvature of the dispersion energy of the rotation phonon with respect to the wave vector $d^2E/dq^2$ gives an idea of the localization of the phonon. A flat dispersion corresponds to a more localized LWF phonon, which will have weaker interactions with its neighbors ($B$), and lower barrier for switching the direction of the rotation. On the opposite, a more dispersive soft-phonon branch gives stronger coupling and more difficult octahedral switching (green line). While this can provide some rough guidelines, there are no easy and general criteria for selecting the materials as the coupling between the octahedral rotation with other lattice distortions can be more complex, and could vary from case to case. Not only the chemistry, but also the pattern of the octahedra rotation (symmetry of the host material) has to be considered. Here we took a prototypical $a^-a^-a^-$ structure but there are others. Preliminary work on some $a^-b^-c^+$ structures shows that they can also be switched.


%
\clearpage
\newpage
\appendix
\onecolumngrid
\renewcommand{\thefigure}{S\arabic{figure}}
\setcounter{figure}{0}
\section*{Supplemental Material}\label{SI}

\subsection{Forces on atoms induced by $\mathrm{Al_{La}}$}
\begin{figure}[htbp]
  \centering
 \includegraphics[width=0.9\textwidth]{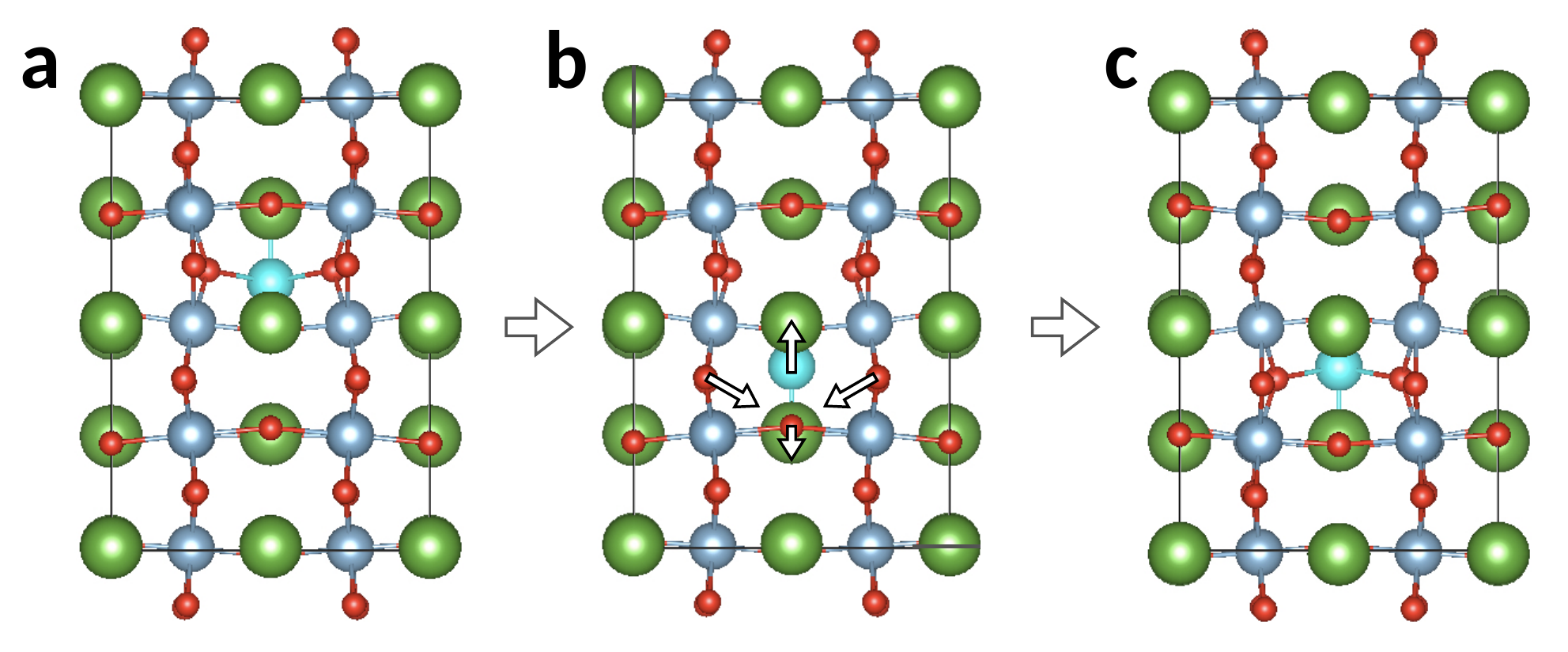}
 \caption{Structure of supercells used to model the defect. The sketches shown in Figure 1a and 1c of the main text correspond to panels (a) and (c), with fully relaxed structures. Panel (b) shows the largest forces on the neighboring O atoms when $\mathrm{Al_{La}}$ is shifted from its ground state position in (a) to a off-centered site along the opposite ({\it wrong}) direction. 
 In this structure, Al$_\mathrm{La}$ attracts the four neighboring O ions on its left and right side, which drives the switching of the octahedral rotations. The white arrows indicate the orientations of the forces on nearby atoms showing that the direction of the octahedral rotations will be switched.  Upon structural relaxation, panel (c) is obtained.
}
\end{figure}

\newpage

\subsection{Map of the staggered amplitudes of the LWF's with $\lambda=1$}

\begin{figure*}[htb]
    \centering
    \includegraphics[width=0.7\textwidth]{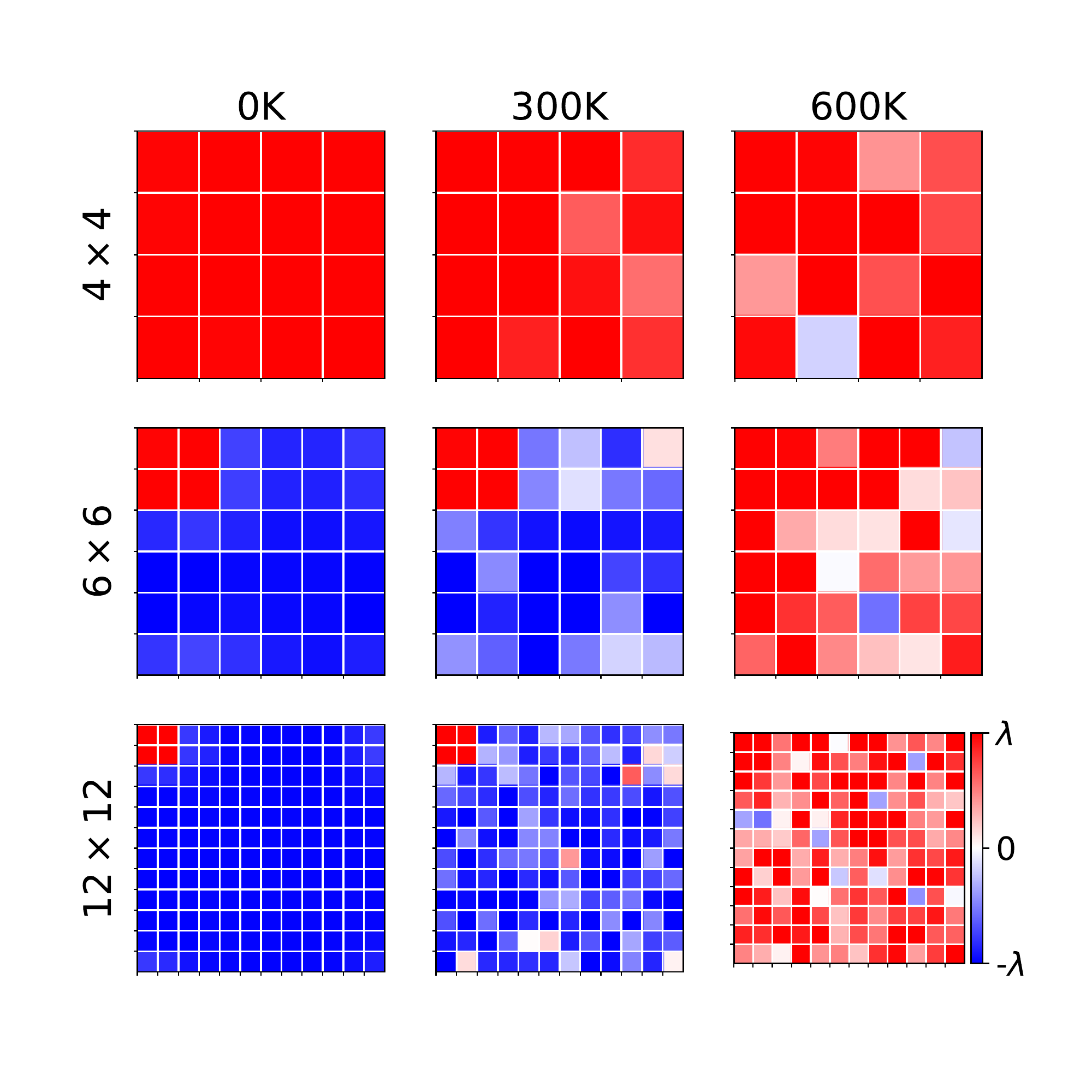}
    \caption{Map of the staggered amplitudes of the LWF's in various supercell size ($4\times 4 \times 4$, $6\times 6 \times 6$, and $12\times 12 \times 12$) from top to bottom, and various temperature (0.1K, 300K, and 600K from left to right) with $\lambda=1$.}
    \label{fig:domain1.0}
\end{figure*}

\clearpage

\subsection{Density of states for the neutral configuration of substitutional Al at the La site (Al$_\mathrm{La}$) in LaAlO$_3$.}

\begin{figure}[h]
\centering
\includegraphics[width=0.55\textwidth]{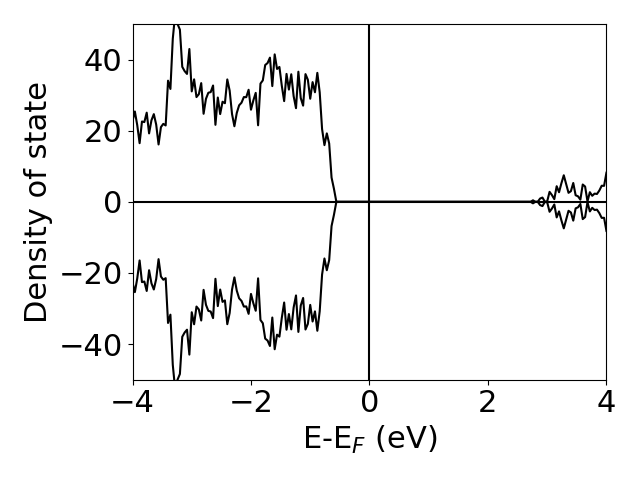}
\label{fig:DOS of neutral LAO-Al}
\caption{Density of states for neutral Al$_\mathrm{La}$ in LaAlO$_3$. The Fermi level is placed in the gap of bulk LAO, and no localized defect levels are visible within the band gap.}
\end{figure}

\newpage

\subsection{Structure and defect level for $\mathrm{Al_{La}^{2-}}$} \label{alla2e}

The relaxed atomic structure for the charged $\mathrm{Al_{La}^{2-}}$ (doped with two extra electrons) is shown in the figure. The substitutional Al recovers the high-symmetry position, with no off-centering. There is no energy barrier in this structural relaxation, and there is a remarkable energy gain of $\sim$1.4 eV. The a$^-$a$^-$a$^-$ tilt pattern is recovered, although the octahedral rotation angles are slightly reduced from the pristine crystal values. The extra electrons occupy a deep defect level in the band gap, which is mostly localized on the Al-$3s$ orbitals, as illustrated in the Density of States. 
Note that the neutral defect is the most stable configuration over a broad range of values of the Fermi level (electronic chemical potential) within the gap, and only when $\mu_e$ is close to the conduction band does the charged defect become more stable (panel c). The formation energies were computed taking the reference values of isolated atoms of Al and La as chemical potentials, and including the FNV correction~\cite{FNV} for charged defects in periodic boundary conditions.

\begin{figure*}[htb]
    \centering
    \includegraphics[width=1.0\textwidth]{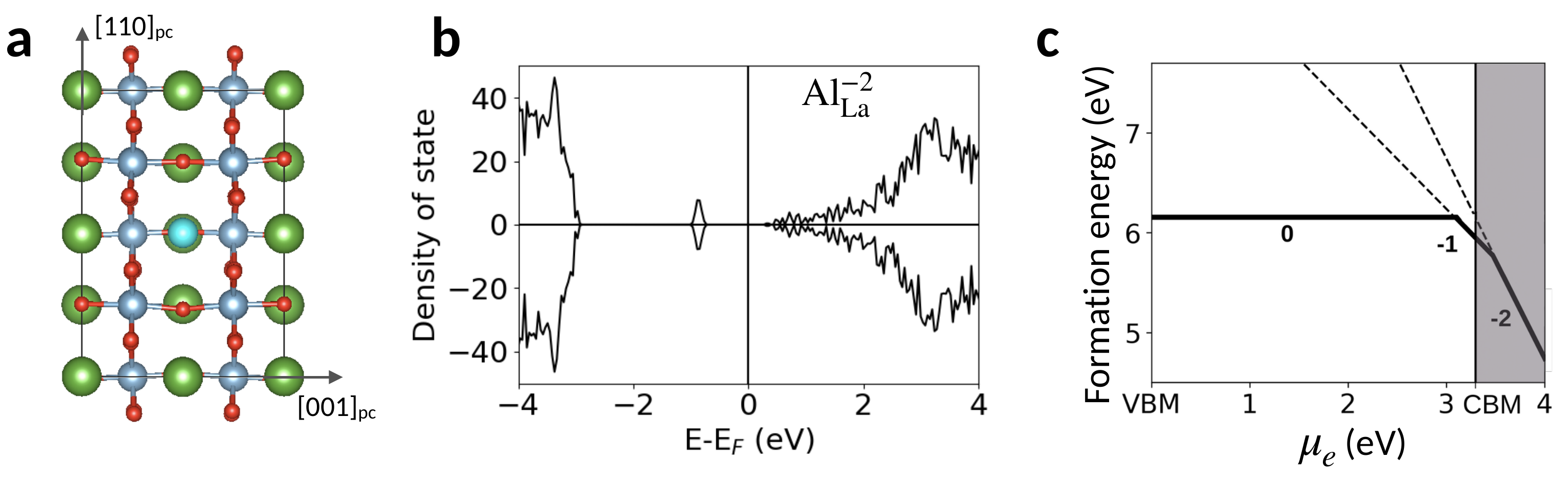}
    \caption{a) Atomic structure visualization for the charged defect, occupying the A-site in the host lattice. b) Density of States showing the occupied defect level in the gap. c) formation energy for Al$_\mathrm{La}^q$ in different charge states (q=0,-1,-2) as a function of the electronic chemical potential between the top of the valence band (VBM) and the bottom of the conduction band (CBM). Thick solid line shows the most stable configuration.}
    \label{fig:rLAO_Al_2e}
    
\end{figure*}


\begin{thebibliography}{78}
\bibitem [{Bor()}]{BornCharges}%
  \BibitemOpen
  \href@noop {} {}\bibinfo {note} {We note that $Z^*(\mathrm{Al_{La}})$ is
  calculated in the structure with defect, and the other Born effective
  charges, $Z^{*}_i$, are taken for the pristine structure to avoid the large
  computational cost. This results in a violation of the sum rule
  \cite{gonze1997dynamical} $\sum_{i=1}^{N} Z^{*}_i=0$, which introduces an
  error in the estimation of the polarization. In the calculation, we force the
  charge center to be fixed, which is equivalent to enforcing the sum rule by
  subtracting the averaged residual Born charge ( $\bar{\text
  Z}=\frac{1}{N}\sum_{i=1}^{N} Z^{*}_i$). For the purpose of a qualitative
  estimation of the polarization and the effectiveness of the electric field on
  the defect movement, this is a good approximation, as the $\bar{\text Z}$ is
  relatively small ($0.014\,e$).}\BibitemShut {Stop}%
\end{thebibliography}
\end{document}